\documentclass[a4paper,10pt]{article}

\pdfoutput=1

\usepackage{rotating}
\usepackage{listings}
\usepackage{amsmath}
\usepackage{amssymb}
\usepackage{amsthm}
\usepackage{mathpartir}
\usepackage{xspace}

\usepackage[utf8]{inputenc}

\usepackage{tikz}
\usepackage{tikz-qtree}

\usepackage[backend=bibtex]{biblatex}

\usepackage{url}
\bibliography{bibliography.bib}

\newlength\digitwidth
\footnotesize
\let\LSTINLINE\lstinline
\def\lstinline{\LSTINLINE[basicstyle=\sffamily\upshape]}

\renewcommand*\thelstnumber{\ifnum\value{lstnumber}<10\relax\kern\digitwidth\fi\arabic{lstnumber}}

\makeatletter
\renewenvironment{figure}
  {\@float{figure}\footnotesize}
  {\end@float}
\makeatother

\let\phi\varphi

\let\variables V

\newcommand*{\mkterm}[1]{\mathsf{#1}}

\newcommand{\reduces}{\longrightarrow}

\newcommand{\bnf}{\;::=\;}
\newcommand{\alt}{\;\;|\;\;}

\usepackage{color}

\newcommand{\ectx}{\mathcal{E}}

\newcommand*{\typedexp}[4]{{#1}\vartriangleright{#2} : {#3}\vartriangleleft{#4}}
\newcommand*{\typedexpc}[4]{{#1}\vartriangleright_{C}{#2} : {#3}\vartriangleleft{#4}}
\def\judgment#1>#2:#3<#4/{\ensuremath{\typedexp{#1}{#2}{#3}{#4}}}

\def\subjudgment#1>#2:#3<#4/{\ensuremath{{#1}\triangleright{#2}\subt{#3}\triangleleft{#4}}}
\def\judgmentc#1>#2:#3<#4/{\ensuremath{\typedexpc{#1}{#2}{#3}{#4}}}

\newcommand{\subt}{\mathrel{\mathsf{<:}}}

\newcommand*{\comment}[1]{}

\newcommand\ignore[1]{}
\makeatletter\let\noqed\@qededtrue\makeatother

\newcommand{\defeq}{\stackrel{\mathrm{def}}{=}}

\lstdefinelanguage{spi}
 {morekeywords=[1]{ok,pair,fst,snd,Un,Key},
  morekeywords=[1]{Ok,Pair,Ch,empty,union},
  morekeywords=[2]{out,in,new,event,if,then,else,let,else,begin,end,exercise,decrypt,as,split,match},
  morekeywords=[3]{TermCtor,TermData,TermDtor,Predicate,Process,Private,TopLevel,Main},
  sensitive=true,%
  morecomment=[l]{//},
  literate={->}{$\to\:$}3,
  literate={<=}{$\leq\:$}3,
  morestring=[b]"}

\title{The While language}
\author{Cláudio Vasconcelos \qquad António Ravara\\\\
  NOVA-LINCS and Dep. de Informática, FCT.\\
  Universidade NOVA de Lisboa, Portugal}

\begin{document}
\maketitle

\begin{abstract}
  This article presents a formalisation of a simple imperative
  programming language.  The objective is to study and develop
  "hands-on" a formal specification of a programming language, namely
  its syntax, operational semantics and type system. To have an
  executable version of the language, we implemented in Racket its operational
  semantics and type system.
\end{abstract}

\section{Introduction}
This article consists on the presentation of the While language
described in a book by Hanne Riis and Flemming
Nielson~\cite{riisnielsonh.nielsonf.2007}. We follow the definitions in
the book, namely to set up the syntax and operational semantics of the
language, and devise a type system, which is thus original.

This is an initial step to 
understand at the same time: (1) the fundamentals of imperative programming and its
features, such as state changes, order of execution and control flow
expressions \cite{ms}; 
and (2) how to have an executable version of the formalisation, allow
to automaticaly build derivations of the reduction semantics and of type-checking.
We decided to do this with a simple imperative programming language
specification before we start studying and modifying more complex
imperative languages, namely languages that include object-oriented
features.

The While language presented in this article is a small imperative
language that allows non-deterministic and parallel execution of
statements and also the use of blocks with local variable and
procedure declarations. The syntax, presented in Section
\ref{sec:while_syntax}, is based on the same syntax of While presented
in chapters two and three of \cite{riisnielsonh.nielsonf.2007},
although we changed some existing constructs and extended it with a
runtime syntax, including a evaluation context.


Section \ref{sec:while_semantics} presents the operational semantics,
based on the one presented in chapter two of
\cite{riisnielsonh.nielsonf.2007}. In addition to those rules, we
defined the reduction rules for non-deterministic, parallel and block
constructs using structural operational semantics because, although
they are presented in chapter three of
\cite{riisnielsonh.nielsonf.2007}, the authors presented them using
only natural semantics. We choose to use structural operational
semantics since small-step reduction allows us to specify in detail
the behaviour of the language in a concurrent context, which it is not
possible using natural semantics.

Section \ref{sec:while_typing} presents the type system we created for
While. In a similar way to the operational semantics, the typing rules
use two different environments, one for declared variables and one for
declared procedures. In the type system, variables are represented by
its type but procedures are represented by a variable environment that
contains all of the variables declared in the procedure. The typing
rules for the expressions receive both environments as input, while
typing rules for statements receive both as input and returns those
environments, possibly modified, as output, so that the next rule is
aware of the changes done to the program state.


Section \ref{sec:racket} presents the Racket language
\cite{manifesto}, a programming language that supports other
programming languages. Racket offers PLT Redex~\cite{redex}, a
domain-specific language embedded in Racket that allows programmers to
formalize and debug programming languages. We implemented the While as
formalized in this article using PLT Redex and tested it using small
programs defined by us. The code of the implementation, along with the
programs we used to test it, is available at
\url{https://bitbucket.org/cvasconcelos/thesis/src/876fc254db76aca1bb058b7e6ef069ee23c6c237/While/while.rkt}.

\section{Syntax}
\label{sec:while_syntax}
For presentation sake, the syntax of While is divided into three parts.

\subsection{Basic syntax}
\label{sec:while_basic_syntax}
The basic syntax of While, defined in Figure \ref{fig:basicsyntax}, is
based on the syntax presented in
\cite[p.~7]{riisnielsonh.nielsonf.2007}, which contains basic
arithmetic and boolean expressions and statements. Let $n$ stand for
an natural value and consider a set of variables symbols ranged over
by $x$. In addition to that, we add some primitive types, a new type
of value for statements ($\mkterm{void}$) and instead of including the
variable assignment expression we created two new statements: One for
variable declaration and one for variable update. Figures
\ref{fig:abstracttree1}, \ref{fig:abstracttree2} and
\ref{fig:abstracttree3} show examples of programs in the abstract
syntax of the While language.
 
 
\subsection{Extended syntax}
\label{sec:while_extended_syntax}
The extended syntax of While, defined in Figure
\ref{fig:extendedsyntax}, is based on some of the advanced constructs
presented in \cite[p.~47 - 56]{riisnielsonh.nielsonf.2007}. Consider a
set of procedure names ranged over by $p$. Blocks and procedures are
added to the syntax, allowing to specify a block inside a program. The
While language has dynamic scope for variables and procedures, meaning
that each block has its own scope.

This extension also specifies $\mkterm{par}$, a construct for parallel
execution of two statements in a interleaved way, and
$\mkterm{protect}$, a construct for atomic execution of a statement.

Figures \ref{fig:abstracttree4} and \ref{fig:abstracttree5} show
examples of programs in the abstract syntax of the While language that
use these new features.

\subsection{Runtime syntax}
\label{sec:while_runtime_syntax}
The runtime syntax of the While language is composed by a set of
constructs necessary during runtime, i.e., by the reduction and/or the
typing rules, and they are not available to the user. The original
While language presented in \cite{riisnielsonh.nielsonf.2007} does not
have any runtime exclusive constructs so we define the necessary
constructs, which are presented in Figure \ref{fig:runtimesyntax}.

A new construct, $\mkterm{protected}$, is added to the statement set
and is used in the operational semantics to help indicating that a
statement must be executed as an atomic entity. This idea is presented
in \cite{protectsdfsdfsdf}.  There is also two new constructs in the
statement set, $\mkterm{beginscope}$ and $\mkterm{endscope}$, that are
used by the operational semantics for scope management. Arithmetic and
boolean expressions and values are also added to the statement set
because during runtime we need to consider them statements for
evaluation purposes.

Finally, the contexts of the While language are defined. These
contexts specify how each expression must be evaluated, more
specifically the order of execution of each expression.

\begin{figure}
  \textbf{Basic Syntax}
  \begin{align*} 
    \text{(Arithmetic expressions)} && a & \bnf n \alt x \\ &&& \quad\alt a + a \alt a - a\alt a * a\\
    \text{(Boolean expressions)} && b & \bnf \mkterm{true} \alt \mkterm{false} \alt x \\ &&& \quad\alt a = a \alt a \le a \alt b \wedge b \\ &&& \quad\alt \neg b\\
    \text{(Values)} && val & \bnf n \alt \mkterm{true} \alt \mkterm{false} \alt \mkterm{void}\\
	\text{(Types)} && t & \bnf \mkterm{Nat} \alt \mkterm{Bool} \alt \mkterm{Cmd}\\
    \text{(Statements)} && S & \bnf S \mkterm{;} S \\ &&& \quad\alt \mkterm{if} \ b \ \mkterm{then} \ S \ \mkterm{else} \ S \alt \mkterm{while} \ b \ \mkterm{do} \ S \\ &&& \quad\alt \mkterm{var} \ t \ x \ := \ a \alt \mkterm{var} \ t \ x \ := \ b \\ &&& \quad\alt x \ := \ a \alt x \ := \ b\\
  \end{align*}
  \caption{Basic syntax}
  \label{fig:basicsyntax}
\end{figure}

\begin{figure}
  \textbf{Extended Syntax}
  \begin{align*}
    \text{(Statements)} && S & \bnf \ldots \alt \mkterm{begin} \ D_v \ D_p \ S \ \mkterm{end} \alt \mkterm{call} \ p \\ &&& \quad\alt S \ \mkterm{par} \ S \alt \mkterm{protect} \ S \ \mkterm{end} \\
    \text{(Variable declarations)} && D_v & \bnf \varepsilon \alt D_v \mkterm{;} D_v \\ &&& \quad\alt \mkterm{var} \ t \ x \ := \ b \alt \mkterm{var} \ t \ x \ := \ a \\
    \text{(Procedure declarations)} && D_p & \bnf \varepsilon \alt D_p \mkterm{;} D_p \alt \mkterm{proc} \ p \ \mkterm{is} \ S
  \end{align*}
  \caption{Extended syntax}
  \label{fig:extendedsyntax}
\end{figure}

\begin{figure}
  \textbf{Runtime Syntax}
  \begin{align*}
    \text{(Statements)} && S & \bnf \ldots\alt a \alt b \alt val \alt \mkterm{beginscope} \alt \mkterm{endscope} \\ &&& \quad\alt \mkterm{protected} \ S \ \mkterm{end}\\      
	\text{(evaluation context)} && \ectx & \bnf n + {\ectx} \alt {\ectx} + a \alt n - {\ectx} \alt {\ectx} - a \\ &&& \quad\alt n * {\ectx} \alt {\ectx} * a \\ &&& \quad\alt n = {\ectx} \alt {\ectx} = a \alt n \le {\ectx} \alt {\ectx} \le a \\ &&& \quad\alt \mkterm{true} \wedge {\ectx} \alt \mkterm{false} \wedge {\ectx} \alt {\ectx} \wedge b \alt \neg {\ectx} \\ &&& \quad\alt {\ectx};S \\ &&& \quad\alt \mkterm{if} \ {\ectx} \ \mkterm{then} \ S_1 \ \mkterm{else} \ S_2 \alt \mkterm{var} \ t \ x \ := \ {\ectx} \alt x \ := \ {\ectx} \\ &&& \quad\alt {\ectx} \ \mkterm{par} \ S \alt S \ \mkterm{par} \ {\ectx} \alt \mkterm{protected} \ {\ectx} \ \mkterm{end}
  \end{align*}
  \caption{Runtime syntax}
  \label{fig:runtimesyntax}
\end{figure}

\begin{figure}
  \textbf{var Nat y := 4; y := y + 1}
  
   \centering
	\begin{tikzpicture}[style={level distance=1.2cm,sibling distance=1.5cm}]
\Tree
[.[Comp] 
  [.[Assign] 
  	[.t Nat ] 
    [.x y ] 
    [.a [.n 4 ]]]
  [.Update 
  	[.x y ] 
    [.+ 
  		[.a [.x y ]] 
    		[.a [.n 1 ]]]]
]
\end{tikzpicture}
  \caption{Abstract syntax example 1}
  \label{fig:abstracttree1}
\end{figure}
\clearpage

\begin{figure}
  \textbf{var Bool y := false; if $\neg$ y then var z := 1 else var Nat z := 3}
	
	\centering
	\begin{tikzpicture}[style={level distance=1.2cm,sibling distance=0.7cm}]
\Tree
[.[Comp] 
  [.[Assign] 
  	[.t Bool ] 
    [.x y ] 
    [.b false  ]] 
  [.If 
	[.[Not] 
  		[.b false ]]
	[.[Assign] 
  		[.t Nat ] 
    		[.x z ] 
    		[.a [.n 1 ]]] 
    [.[Assign] 
  		[.t Nat ] 
    		[.x z ] 
    		[.a [.n 3 ]]]]
]
\end{tikzpicture}
  \caption{Abstract syntax example 2}
  \label{fig:abstracttree2}
\end{figure}

\begin{figure}
  \textbf{var Nat y := 0; while y = 0 do y := y + 1}
	
	\centering
	
\begin{tikzpicture}[style={level distance=1.2cm,sibling distance=1.2cm}]
\Tree
[.[Comp] 
  [.[Assign] 
  	[.t Nat ] 
    [.x y ] 
    [.a [.n 0 ]]] 
  [.[While] 
	[.$\le$ 
  		[.a [.x y ]]
  		[.a [.n 0 ]]]
	[.[Update] 
  	[.x y ] 
    [.+ 
  		[.a [.x y ]] 
    		[.a [.n 1 ]]]]
]]
\end{tikzpicture}
  \caption{Abstract syntax example 3}
  \label{fig:abstracttree3}
\end{figure}

\begin{figure}
\textbf{begin var w := 2; proc z is var Nat r = 4; call z; w := r end}
	
	\centering
	
	\begin{tikzpicture}[style={level distance=1.2cm,sibling distance=0.6cm}]
\Tree
[.[Begin] 
  [.[Assign] 
  	[.t Nat ] 
    [.x w ] 
    [.a [.n 2 ]]] 
  [.[Proc] 
  	[.p z ] 
    [.S [.[Assign] 
  	[.t Nat ] 
    [.x b ] 
    [.a [.n 4 ]]  ]] 
  [.[Seq] 
	[.[Call] 
  		[.p z ]] 
    [.[Update] 
  	[.x a ] 
    [.a [.x b ]]  ]] ]
]]
\end{tikzpicture}

  \caption{Abstract syntax example 4}
  \label{fig:abstracttree4}
\end{figure}

\begin{figure}
\textbf{var Nat x := 0; protect x := 2; x := 4 end par x := 6}
	
	\centering
	
	\begin{tikzpicture}[style={level distance=1.2cm,sibling distance=1cm}]
\Tree
[.[Seq] 
  [.[Assign] 
  	[.t Nat ] 
    [.x x ] 
    [.a [.n 0 ]]] 
  [.[Par] 
  	[.[Protect] 
  			[.[Seq] 
  				[.[Update]  
    					[.x y ] 
    					[.a [.n 2 ]]] 
    				[.[Update] 
    					[.x y ] 
    					[.a [.n 4 ]]]]]
  	  [.[Update] 
  		[.x x ] 
    		[.a [.n 6 ]]  ]] ]
]]
\end{tikzpicture}

  \caption{Abstract syntax example 5}
  \label{fig:abstracttree5}
\end{figure}

\section{Operational semantics}
\label{sec:while_semantics}
The operational semantics of While uses a structural operational
semantics approach to specify the behavior of any program in
While. The relation is rigorously defined by a set of reduction
rules. Most of these rules, more specifically the ones for the basic
syntax related expressions, are based on the reduction rules presented
in \cite[p.~33 - 35]{riisnielsonh.nielsonf.2007}. The form of the
rules (judgments)
is 

\begin{mathpar}
\inferrule*[right = $n \geq 0$] 
    { P_1 \\ ... \\ P_n} 
    {\sigma \ \rho \ \vdash \ S \reduces S \ \dashv \ \sigma \ \rho} 
   \end{mathpar} 

An element of the relation is a pair of triples ($\sigma, \rho, S $) where $\sigma$ and $\rho$ are two environments and $S$ is a statement (c.p. figure \ref{fig:basicsyntax}). Here we consider an environment to be a sequence of maps. In $\sigma$ each variable \textit{x} is mapped to a value \textit{val}, while in $\rho$ each procedure \textit{p} is mapped to a statement \textit{S}. Both of this environments consist of several "levels", each one represented by a map with all of the variables or procedures that belong to a program block it represents. Each environment starts with one level, which is always the global scope, while the "levels" created afterwards represent the scope of program blocks, with the earliest "level" representing the first program block, the "level" after representing the program block inside the first program block, and so on. Each time we get inside a program block a new "level" is added to both environments, and this "level" will be removed when we get out of that same program block. Section \ref{sec:reducexample} shows a reduction of a simple program with a program block where this is shown.

The relation is inductively defined by the rules in Figures \ref{fig:reduc1}, \ref{fig:reduc2} and \ref{fig:reduc3}. The semantics for arithmetic and boolean expressions is omitted because it is identical to the one presented in \cite[p.~13 - 15]{riisnielsonh.nielsonf.2007}.

Figure \ref{fig:reduc1} presents the reductions rules for basic statements. Rules \textsc{Assign} and \textsc{Update} are both for variable assignment, with the first one being for a new variable declaration, which maps a new variable \textit{x} to value \textit{val} in $\sigma$ , and the second one being for variable update, which updates the value of \textit{x} to the new value \textit{$val_2$}. Rules \textsc{Seq1} and \textsc{Seq2} are for sequential compositions. 

Since the operational semantics of While is based on a structural operational semantics approach, in a sequential composition the statement $S_1$ does not necessarily terminate on one computational step. The rule \textsc{Seq1} expresses this situation, while rule \textsc{Seq2} expresses a situation where $S_1$ completely terminates.

Rules \textsc{If-True} and \textsc{If-False} are for $\mkterm{if-then-else}$ expressions, reducing them to one of its branches depending on the boolean value that serves as the condition. The condition is evaluated using the semantics of boolean expressions in \cite[p.~15]{riisnielsonh.nielsonf.2007}. Axiom \textsc{While} unfolds $\mkterm{while-do}$ expressions into $\mkterm{if-then-else}$ expressions, with the first branch being the execution of statement \textit{S} and then the execution of the same while expression again, and the second one being a empty branch. 

Figure \ref{fig:reduc2} presents the reduction rules for block and procedure related statements. Rule \textsc{Begin} creates a sequence of statements that, in a new scope, will assign variables and procedures and execute the body of the block. Axioms \textsc{BeginScope} and \textsc{EndScope} are for scope management, with the first being used to create a new "level" on the top of both variable environments and the second one to destroy those same "levels". 

\textsc{Proc} maps a new variable \textit{x} to a statement \textit{S} in $\rho$ while \textsc{Call} reduces a variable \textit{x} to a statement \textit{S} to which it is mapped in $\rho$.

Figure \ref{fig:reduc3} shows the reduction rules for the parallelism and concurrency related statements \textsc{Par1}, \textsc{Par2}, \textsc{Par3} and \textsc{Par4} are for parallel execution of statements and they reflect the non deterministic and interleaved execution of the statements. The predicate $protected$ used in this four rules is presented in \cite{protectsdfsdfsdf}. Rule \textsc{Protect} works similar to a lock mechanism, where a statement \textit{S} obtains a lock if available so it can execute as an atomic entity. Rule \textsc{Protected} allows that same statement to release the lock.

\begin{figure}
  \centering
  \begin{mathpar}
  \inferrule*[left=Assign \ ]
    {|\sigma' | = 1 \wedge x \notin \text{dom}(\sigma' )}
  {(\sigma , \sigma')  \ \rho \ \vdash \ \mkterm{var} \ t \ x := val \reduces \mkterm{void} \ \dashv \ (\sigma , \sigma' \cup  \{x \mapsto val\}) \ \rho}
  
  \inferrule*[left=Update \ ]
    {x \in \text{dom}(\sigma )}
  {\sigma\{x \mapsto val_1\} \ \rho \ \vdash \ x := val_2 \reduces \mkterm{void} \ \dashv \ \sigma\{x \mapsto val_2\} \ \rho}
	
	\inferrule*[left=Seq1 \ ]
    {\sigma \ \rho \ \vdash \ S_1 \reduces S_1' \ \dashv \ \sigma' \ \rho'}
    {\sigma \ \rho \ \vdash \ S_1;S_2 \reduces S_1';S_2 \ \dashv \ \sigma' \ \rho'}  
  
  \inferrule*[left=Seq2 \ ]
    {\sigma \ \rho \ \vdash \ S_1 \reduces \mkterm{void} \ \dashv \ \sigma' \ \rho'}
    {\sigma \ \rho \ \vdash \ S_1;S_2 \reduces S_2 \ \dashv \ \sigma' \ \rho'}  
    
    \inferrule*[left=If-True \ ]
    { b = \mkterm{true}}
    {\ \ \sigma \ \rho \ \vdash \ \mkterm{if} \ b \ \mkterm{then} \ S_1 \ \mkterm{else} \ S_2 \ \reduces S_1 \dashv \ \sigma \ \rho}  
    
    \inferrule*[left=If-False \ ]
    { b = \mkterm{false}}
    {\ \ \sigma \ \rho \ \vdash \ \mkterm{if} \ b \ \mkterm{then} \ S_1 \ \mkterm{else} \ S_2 \ \reduces S_2 \dashv \ \sigma \ \rho}  
    
    \inferrule*[left=While \ ]
    {}
    {\ \ \sigma \ \rho \ \vdash \ \mkterm{while} \ b \ \mkterm{do} \ S \reduces \ \mkterm{if} \ b \ \mkterm{then} \ S; \ \mkterm{while} \ b \ \mkterm{do} \ S \ \mkterm{else} \ \mkterm{void} \dashv \ \sigma \ \rho}  
  \end{mathpar}
  \caption{Reduction rules for basic statements}
  \label{fig:reduc1}
\end{figure}

\begin{figure}
  \centering
  \begin{mathpar}
  \inferrule*[left=Begin \ ]
    {}
    {\ \ \sigma \ \rho \ \vdash \ \mkterm{begin} \ D_v \ D_p \ S \reduces \ \mkterm{beginscope};D_v;D_p;S;\mkterm{endscope} \dashv \ \sigma' \ \rho}  
  
  \inferrule*[left=BeginScope \ ]
    {}
  {\sigma \ \rho \ \vdash \ \mkterm{beginscope} \reduces \mkterm{void} \dashv \ (\sigma' , \sigma) \ (\rho' , \rho)}
  
    \inferrule*[left=EndScope \ ]
    {}
  {(\sigma' , \sigma) \ (\rho' , \rho) \ \vdash \ \mkterm{endscope} \reduces \mkterm{void} \dashv \ \sigma \ \rho}
  
    \inferrule*[left=Proc \ ]
    {|\rho' | = 1 \wedge p \notin \text{dom}(\rho' )}
  {\sigma \ (\rho , \rho') \ \vdash \ \mkterm{proc} \ p \ \mkterm{is} \ S \reduces void \dashv \ \sigma \ (\rho , \rho' \cup \{p \mapsto S\})}
  
  \inferrule*[left=Call \ ]
    {p \in \text{dom}(\rho' )}
  {\sigma \ \rho\{p \mapsto S\} \ \vdash \ \mkterm{call} \ p \reduces S \dashv \ \sigma' \ \rho\{p \mapsto S\}}
  \end{mathpar}
  \caption{Reduction rules for blocks and procedures statements}
  \label{fig:reduc2}
\end{figure}

\begin{figure}
  \centering
  \begin{mathpar}
  \inferrule*[left=Par1 \ ]
    {\sigma \ \rho \ \vdash \ S_1 \reduces S_1' \ \dashv \ \sigma' \ \rho' \\ \neg \text{protected}(S_1)}
  {\sigma \ \rho \ \vdash \ S_1 \ \mkterm{par} \ S_2 \reduces S_1' \ par \ S_2 \ \dashv \ \sigma' \ \rho'}
  
  \inferrule*[left=Par2 \ ]
    {\sigma \ \rho \ \vdash \ S_1 \reduces void \ \dashv \ \sigma' \ \rho' \\ \neg \text{protected}(S_1)}
  {\sigma \ \rho \ \vdash \ S_1 \ \mkterm{par} \ S_2 \reduces S_2 \ \dashv \ \sigma' \ \rho'}
  
  \inferrule*[left=Par3 \ ]
    {\sigma \ \rho \ \vdash \ S_2 \reduces S_2' \ \dashv \ \sigma' \ \rho' \\ \neg \text{protected}(S_2)}
  {\sigma \ \rho \ \vdash \ S_1 \ \mkterm{par} \ S_2 \reduces S_1 \ par \ S_2' \ \dashv \ \sigma' \ \rho'}
  
  \inferrule*[left=Par4 \ ]
    {\sigma \ \rho \ \vdash \ S_2 \reduces S_2' \ \dashv \ \sigma' \ \rho' \\ \neg \text{protected}(S_2)}
  {\sigma \ \rho \ \vdash \ S_1 \ \mkterm{par} \ S_2 \reduces S_1 \ \dashv \ \sigma' \ \rho'}
    
     \inferrule*[left=Protect \ ]
    {}
  {\sigma \ \rho \ \vdash \ \mkterm{protect} \ S \ \mkterm{end} \reduces \mkterm{protected} \ S \ \mkterm{end} \ \dashv \ \sigma \ \rho}

\inferrule*[left=Protected \ ]
    {}
  {\sigma \ \rho \ \vdash \ \mkterm{protected} \ val \ \mkterm{end} \reduces \mkterm{void} \ \dashv \ \sigma \ \rho}
  \end{mathpar}
  \caption{Reduction rules for parallelism and concurrency statements}
  \label{fig:reduc3}
\end{figure}

\begin{figure}
    \begin{align*}
    protected(S) & \defeq
                       \begin{cases}
                         tt  & \text{ if } S \ \text{is} \ \mkterm{protect} \ S \ \mkterm{end}\\
                         protected(S_1)  & \text{ if } S \ \text{is} \ S_1 ; S_2\\
                         protected(S_1) \vee protected(S_2)  & \text{ if } S \ is \ S_1 \ \mkterm{par} \ S_2\\
                         \mathit{ff} 	& \text{otherwise }                        
                       \end{cases}
\end{align*}
  \caption{Protected predicate}
  \label{fig:protected}
\end{figure}

\clearpage 

\subsection{Reduction example}
\label{sec:reducexample}
Consider the program 

\begin{mathpar}
  \inferrule*
  {}  
  {\mkterm{begin} \ \mkterm{var} \ \mkterm{Nat} \ a := 4; \ b := 2}
\end{mathpar}

and the environments $\sigma = ( \{a \mapsto 3, b \mapsto 5\} )$ and  $\rho = ( \{ \ \} )$. First, we apply the axiom \textsc{Begin} and get

\begin{mathpar}
  \inferrule*
  {}  
  {\sigma \ \rho \ \vdash \ \mkterm{begin} \ \mkterm{var} \ \mkterm{Nat} \ a := 4; \ b := 2 \reduces S \dashv \ \sigma \ \rho}
\end{mathpar}
  
where $S = \mkterm{beginscope}; \ \mkterm{var} \ \mkterm{Nat} \ a := 4; \ b :=  2; \ \mkterm{endscope}$. Using the rule \textsc{Seq1} and the axiom \textsc{BeginScope} we get 

\begin{mathpar}
  \inferrule*
  {\sigma \ \rho \ \vdash \ \mkterm{beginscope} \reduces \mkterm{void} \dashv \ \sigma^{1} \ \rho^{1}}
  {\sigma \ \rho \ \vdash \ S \reduces S' \dashv \ \sigma^{1} \ \rho^{1}}
\end{mathpar}
  
where  $\sigma^{1} = ( \{a \mapsto 3, b \mapsto 5\}, \{ \ \} )$,  $\rho = ( \{ \ \}, \{ \ \} )$ and $S'$ = $\mkterm{var} \ \mkterm{Nat} \ a := 4; \ b := 2; \ \mkterm{endscope}$. Notice that both environments now have a new "level". Using the rules \textsc{Seq1} and \textsc{Assign} we get

 \begin{mathpar}
  \inferrule*
  {\inferrule*
  {x \notin \sigma^{1}}
  {\sigma^{1} \ \rho^{1} \ \vdash \  \mkterm{var} \ \mkterm{Nat} \ a := 4 \reduces \mkterm{void} \dashv \ \sigma^{2} \ \rho^{1}}}
  {\sigma^{1} \ \rho^{1} \ \vdash \ S' \reduces b := 2; \ \mkterm{endscope} \dashv \ \sigma^{2} \ \rho^{1}}
\end{mathpar}

where$\sigma^{2} = ( \{a \mapsto 3, b \mapsto 5\}, \{ a \mapsto 4 \} )$. Although $a$ exists in $\sigma^{2}$, the rule \textsc{Assign} checks if $a$ exists in the deepest "level" of $\sigma^{1}$ which is the case. So, $a$ will be added to the deepest "level" and will be mapped to value $4$. Continuing the reduction process, we now use the rules \textsc{Seq1} and \textsc{Update} to get

 \begin{mathpar}
  \inferrule*
  {\inferrule*
  {x \in \sigma'}
  {\sigma^{2} \ \rho^{1} \ \vdash \  b := 2 \reduces \mkterm{void} \dashv \ \sigma^{3} \ \rho^{1}}}
  {\sigma^{2} \ \rho^{1} \ \vdash \ b := 2; \ \mkterm{endscope} \reduces \mkterm{endscope} \dashv \ \sigma^{3} \ \rho^{1}}
\end{mathpar}

where  $\sigma^{3} = ( \{a \mapsto 3, b \mapsto 2\}, \{ a \mapsto 4 \} )$. The rule \textsc{Update} updated the value of the most recent mapping of $b$, which is the only one in the first "level". Finally, we use the rule \textsc{EndScope} and get

 \begin{mathpar}
  \inferrule*
  {}
  {\sigma^{3} \ \rho \ \vdash \ \mkterm{endscope} \reduces \mkterm{void} \dashv \ \sigma^{4} \ \rho}
\end{mathpar}

where $\sigma^{4} = ( \{a \mapsto 3, b \mapsto 2\} )$. Notice that both the environments now had their last level, both corresponding to the scope of the block we just terminated, removed, while the changes made in other "levels" inside the block still remain. We reached the end of the program.

\section{Type system}
\label{sec:while_typing}
Since \cite{riisnielsonh.nielsonf.2007} does not present a type system for While, we define one from scratch. This type system, similarly to the operational semantics, uses two typing environments: $\Gamma$ for variables and $\Delta$ for procedures. In $\Gamma$ each variable \textit{x} is mapped to a type \textit{t}, and in $\Delta$ each procedure \textit{p} is mapped to a typing environment $\Gamma$ with all the variables declared inside that procedure. In this context we consider an environment to be just one map instead of several maps.

The typing rules of this type system for expressions have the following form:

\begin{mathpar}
\inferrule*[right = $n \geq 0$]
    {P_1 \\ ... \\ P_n}
  {\Gamma \ \Delta \ \vdash \ S : t}
\end{mathpar}

The typing rules for commands have a similar form to the reduction rules, with input and output environments:

\begin{mathpar}
\inferrule*[right = $n \geq 0$]
    {P_1 \\ ... \\ P_n}
  {\Gamma \ \Delta \ \vdash \ S : t \ \dashv \ \Gamma \ \Delta}
\end{mathpar}

Figures \ref{fig:typing1} to \ref{fig:typing4} show the typing rules of While.

Figure \ref{fig:typing1} shows the typing axioms for values. Axioms \textsc{T-True}, \textsc{T-False}, \textsc{T-Nat} and \textsc{T-Void} just evaluate simple values, while axiom \textsc{T-Var} evaluates a variable \textit{x} based on its most recent mapping in $\Gamma$.  

Figure \ref{fig:typing2} shows the typing rules for all arithmetic and boolean expressions of While. In each of this rules the type checker evaluates the expression, checking if each operand has the correct type for the expression. 

Figure \ref{fig:typing3} has the typing rules for simple statements of While. Rule \textsc{T-Assign} evaluates first a statement \textit{S} and then maps a variable \textit{x} to the type \textit{t} of \textit{S}. Rule \textsc{T-Update} just checks if the statement \textit{S} has the same type has the variable to be updated. 

Rule \textsc{T-Seq} evaluates the first statement and then evaluates the second statement taking in consideration all the changes caused by the first statement. 

Rule \textsc{T-If} checks if the expressions that serves has the condition is of type Bool, evaluates both branches in the same conditions (same typing environments as input) and returns a new $\Gamma$ which is the union between both $\Gamma$ returned by each branch. Rule \textsc{T-While} also checks the condition first but it specifies that \textit{S} must not change any of the typing environments (so it does not allow variable assignment in \textit{S}).

Figure \ref{fig:typing4} shows the typing rules for blocks and procedures statements: Rule \textsc{T-Begin} evaluates $D_v$, $D_p$ and $S$ in a similar to rule \textsc{T-Seq}, with the environments returned by each statement to be used as the input typing environments for the next statement. In the end, this rule returns the same typing environments used as input since every change done to these environments can only be visible inside the block.

For procedures, rule \textsc{T-Proc} evaluates a statement \textit{S} and maps a procedure \textit{p} to the typing environment $\Gamma$ returned by \textit{S} in $\Delta$, and rule \textsc{T-Call} returns as $\Gamma$ the union of the $\Gamma$ given as input and the $\Gamma$ which \textit{p} is mapped to.

Figure \ref{fig:typing5} shows the typing rules for the concurrent and parallel statements \textsc{T-Par} and \textsc{T-Protect}. Both just evaluate their statements \textit{S}. 

\begin{figure}
  \centering
  \begin{mathpar}
  \inferrule*[left=T-Nat \ ]
    {}
  {\Gamma \ \Delta \ \vdash \ n : \mkterm{Nat}}
  
 \inferrule*[left=T-Var \ ]
    {}
 	{\Gamma\{x \mapsto t\} \ \Delta \ \vdash \ x : t} 
 	
    \inferrule*[left=T-True \ ]
    {}
  {\Gamma \ \Delta \ \vdash \ \mkterm{true} : \mkterm{Bool}}
  
    \inferrule*[left=T-False \ ]
    {}
  {\Gamma \ \Delta \ \vdash \ \mkterm{false} : \mkterm{Bool}}
  
  \inferrule*[left=T-Empty \ ]
    {}
 	{\Gamma \ \Delta \ \vdash \ \epsilon : \mkterm{Cmd}} 
  \end{mathpar}
  \caption{Typing rules for values}
  \label{fig:typing1}
\end{figure}

\begin{figure}
  \centering
  \begin{mathpar}
  \inferrule*[left=T-Add \ ]
    {\Gamma \ \Delta \ \vdash \ a_1 : \mkterm{Nat} \\ \Gamma \ \Delta \ \vdash \ a_2 : \mkterm{Nat}}
  {\Gamma \ \Delta \ \vdash \ a_1 + a_2 : \mkterm{Nat}}

  \inferrule*[left=T-Sub \ ]
    {\Gamma \ \Delta \ \vdash \ a_1 : \mkterm{Nat} \\ \Gamma \ \Delta \ \vdash \ a_2 : \mkterm{Nat}}
  {\Gamma \ \Delta \ \vdash \ a_1 - a_2 : \mkterm{Nat}}
  
  \inferrule*[left=T-Mult \ ]
    {\Gamma \ \Delta \ \vdash \ a_1 : Nat \\ \Gamma \ \Delta \ \vdash \ a_2 : \mkterm{Nat}}
  {\Gamma \ \Delta \ \vdash \ a_1 * a_2 : \mkterm{Nat}}

  \inferrule*[left=T-Equal \ ]
    {\Gamma \ \Delta \ \vdash \ a_1 : \mkterm{Nat} \\ \Gamma \ \Delta \ \vdash \ a_2 : \mkterm{Nat}}
  {\Gamma \ \Delta \ \vdash \ a_1 = a_2 : \mkterm{Bool}}
  
  \inferrule*[left=T-LEqual \ ]
    {\Gamma \ \Delta \ \vdash \ a_1 : \mkterm{Nat} \\ \Gamma \ \Delta \ \vdash \ a_2 : \mkterm{Nat}}
  {\Gamma \ \Delta \ \vdash \ a_1 \le a_2 : \mkterm{Bool} \ \dashv \ \Gamma \ \Delta}
  
  \inferrule*[left=T-LEqual \ ]
    {\Gamma \ \Delta \ \vdash \ b_1 : \mkterm{Bool} \\ \Gamma \ \Delta \ \vdash \ b_2 : \mkterm{Bool}}
  {\Gamma \ \Delta \ \vdash \ b_1 \wedge b_2 : \mkterm{Bool}}
    
    \inferrule*[left=T-Not \ ]
    {\Gamma \ \Delta \ \vdash \ b : \mkterm{Bool}}
  {\Gamma \ \Delta \ \vdash \ \neg \ b : \mkterm{Bool}}
  \end{mathpar}
  \caption{Typing rules for arithmetic and boolean expressions}
  \label{fig:typing2}
\end{figure}

\begin{figure}
  \centering
  \begin{mathpar}
  \inferrule*[left=T-Assign \ ]
    {\Gamma \ \Delta \ \vdash \ S : t \ \dashv \ \Gamma \ \Delta}
  {\Gamma \ \Delta \ \vdash \ \mkterm{var} \ t \ x := e : \mkterm{Cmd} \ \dashv \ \Gamma \cup \{x \mapsto t\} \ \Delta}
  
  \inferrule*[left=T-Update \ ]
    {\Gamma \ \Delta \ \vdash \ S : t \ \dashv \ \Gamma \ \Delta}
  {\Gamma\{x \mapsto t\} \ \Delta \ \vdash \ x := S : \mkterm{Cmd} \ \dashv \ \Gamma\{x \mapsto t\} \ \Delta}

  \inferrule*[left=T-Seq \ ]
    {\Gamma \ \Delta \ \vdash \ S_1 : t_1 \ \dashv \ \Gamma' \ \Delta \\ \Gamma' \ \Delta \ \vdash \ S_2 : t_2 \ \dashv \ \Gamma'' \ \Delta}
  {\Gamma \ \Delta \ \vdash \ S_1 ; S_2 : t_2 \ \dashv \ \Gamma'' \ \Delta}

  \inferrule*[left=T-If \ ]
    {\Gamma \ \Delta \ \vdash \ b : \mkterm{Bool} \\ \Gamma \ \Delta \ \vdash \ S_1 : t \ \dashv \ \Gamma' \ \Delta \\ \Gamma \ \Delta \ \vdash \ S_2 : t \ \dashv \ \Gamma'' \ \Delta}
  {\Gamma \ \Delta \ \vdash \ \mkterm{if} \ b \ \mkterm{then} \ S_1 \ \mkterm{else} \ S_2 : t \ \dashv \ (\Gamma'' \cup \Gamma'') \ \Delta}

   \inferrule*[left=T-While \ ]
    {\Gamma \ \Delta \ \vdash \ b : \mkterm{Bool} \\ \Gamma \ \Delta \ \vdash \ S : \mkterm{Cmd} \ \dashv \ \Gamma \ \Delta}
  {\Gamma \ \Delta \ \vdash \ \mkterm{while} \ b \ \mkterm{do} \ S : \mkterm{Cmd} \ \dashv \ \Gamma \ \Delta}
  \end{mathpar}
  \caption{Typing rules for simple statements}
  \label{fig:typing3}
\end{figure}

\begin{figure}
  \centering
  \begin{mathpar}  
  \inferrule*[left=T-Begin \ ]
    {\Gamma \ \Delta \ \vdash \ D_v : \mkterm{Cmd} \ \dashv \ \Gamma' \ \Delta \\ \Gamma' \ \Delta \ \vdash \ D_p : \mkterm{Cmd} \ \dashv \ \Gamma' \ \Delta' \\ \Gamma' \ \Delta' \ \vdash \ S : \mkterm{Cmd} \ \dashv \ \Gamma'' \ \Delta}
  {\Gamma \ \Delta \ \vdash \ \mkterm{begin} \ D_v \ D_p \ S \ \mkterm{end} : \mkterm{Cmd} \ \dashv \ \Gamma \ \Delta}  

  \inferrule*[left=T-Call \ ]
    {}
  {\Gamma \ \Delta\{p \mapsto \Gamma'\} \ \vdash \ \mkterm{call} \ p : \mkterm{Cmd} \ \dashv \ (\Gamma \cup \Gamma') \ \Delta\{p \mapsto \Gamma'\}}

  \inferrule*[left=T-Proc \ ]
    {\Gamma \ \Delta \ \vdash \ S : \mkterm{Cmd} \ \dashv \ \Gamma' \ \Delta}
  {\Gamma \ \Delta \ \vdash \ \mkterm{proc} \ p \ \mkterm{is} \ S  : \mkterm{Cmd} \ \dashv \ \Gamma \ \Delta \cup \{p \mapsto \Gamma' \backslash \Gamma \}}
  \end{mathpar}
  \caption{Typing rules for block statements}
  \label{fig:typing4}
\end{figure}

\begin{figure}
  \centering
  \begin{mathpar}  
  \inferrule*[left=T-Par \ ]
    {\Gamma \ \Delta \ \vdash \ S_1 : t_1 \ \dashv \ \Gamma' \ \Delta \\ \Gamma \ \Delta \ \vdash \ S_2 : t_2 \ \dashv \ \Gamma'' \ \Delta}
  {\Gamma \ \Delta \ \vdash \ S_1 \ \mkterm{par} \ S_2 : \mkterm{Cmd} \ \dashv \ (\Gamma' \cup \Gamma'') \ \Delta}
  
  \inferrule*[left=T-Protect \ ]
    {\Gamma \ \Delta \ \vdash \ S : t \ \dashv \ \Gamma' \ \Delta}
  {\Gamma \ \Delta \ \vdash \ \mkterm{protect} \ S \ \mkterm{end} : \mkterm{Cmd} \ \dashv \ \Gamma' \ \Delta}
  \end{mathpar}
  \caption{Typing rules for concurrent statements}
  \label{fig:typing5}
\end{figure}

Figures \ref{fig:typingexample1}, \ref{fig:typingexample2} and \ref{fig:typingexample3} show examples of derivations of correct programs in While using the type system defined. Figure  \ref{fig:typingexample4} shows an example of a badly constructed program that the type checker fails to evaluate due to the scope defined for While. In this example, when applying the rule \textsc{T-Add}, the type checker expects \textit{y} to be of type Nat and while there is one \textit{y} of type Nat, it is inside a block, so the current \textit{y} is of type Bool.

\begin{figure}
  \centering
  \begin{mathpar}  
  \inferrule*[right=T-If] {
	\inferrule*[right=T-Not]
    		{
    			\inferrule*[right=T-True]
    				{ }
	  			{\Gamma \ \Delta \ \vdash \ \mkterm{true} : \mkterm{Bool} \ \dashv \ \Gamma \ \Delta}    
  		}
  		{\Gamma \ \Delta \ \vdash \ \neg \ \mkterm{true} : t \ \dashv \ \Gamma \ \Delta}    
  	\\ \text{T1} \\ \text{T2}
  }
  {\Gamma \ \Delta \ \vdash \ \mkterm{if} \ \neg \mkterm{true} \ \mkterm{then} \ \mkterm{var} \ \mkterm{Nat} \ y := 2 \ \mkterm{else} \ \mkterm{var} \ \mkterm{Nat} \ z := 4 : t \ \dashv \ \Gamma\{y \mapsto \mkterm{Nat}, z \mapsto \mkterm{Nat}\} \ \Delta}
  
  \\
  
  \inferrule*[left=T1 \ ,right= T-Assign ]
    		{
    			\inferrule*[right= T-Nat]
    				{ }
  				{\Gamma \ \Delta \ \vdash \ 2 : \mkterm{Nat} \ \dashv \ \Gamma \ \Delta}  
  		}
  		{\Gamma \ \Delta \ \vdash \ \mkterm{var} \ \mkterm{Nat} \ y := 2 : \mkterm{Cmd} \ \dashv \ \Gamma\{y \mapsto \mkterm{Nat}\} \ \Delta}  
  		
  		\\
  		
  	\inferrule*[left=T2 \ ,right= T-Assign]
    		{
    			\inferrule*[right= T-Nat]
    				{ }
  				{\Gamma \ \Delta \ \vdash \ 4 : \mkterm{Nat} \ \dashv \ \Gamma \ \Delta}  
  		}
  		{\Gamma \ \Delta \ \vdash \ \mkterm{var} \ \mkterm{Nat} z := 4 : \mkterm{Cmd} \ \dashv \ \Gamma\{z \mapsto \mkterm{Nat}\} \ \Delta}  
  		
  \end{mathpar}
  \begin{flushleft}
  $\Gamma = \varnothing$ \\
  $\Gamma^{1} = \{y \mapsto \mkterm{Nat}\}$ \\
  $\Gamma^{2} = \{z \mapsto \mkterm{Nat}\}$ \\
  $\Gamma^{3} = \{y \mapsto \mkterm{Nat},z \mapsto \mkterm{Nat}\}$\\
  $\Delta = \varnothing$
  \end{flushleft}

  \caption{Typing example 1}
  \label{fig:typingexample1}
\end{figure}

\begin{figure}
  \centering
  \begin{mathpar}  
  \inferrule*[right=T-Seq \ ]
  { 
	\inferrule*[right= T-Assign\ ]
    		{
    			\inferrule*[right= T-Nat \ ]
    				{ }
  				{\Gamma \ \Delta \ \vdash \ 1 : \mkterm{Nat} \ \dashv \ \Gamma \ \Delta}  
  		}
  		{\Gamma \ \Delta \ \vdash \ \mkterm{var} \ \mkterm{Nat} \ x := 4 : \mkterm{Cmd} \ \dashv \ \Gamma^{1} \ \Delta}
  		\\ T1
    }
	  {\Gamma \ \Delta \ \vdash \ \mkterm{var} \ \mkterm{Nat} \ x := 1 ; \mkterm{while} \ x \le 4 \ \mkterm{do} \ x := x + 1 : \mkterm{Cmd} \ \dashv \ \Gamma^{1} \ \Delta}    
  		
  		\\
  		
  		\inferrule*[left=T1 \ ,right= T-While\ ]
    		{
    			\inferrule*[right= T-LEqual \ ]
    				{ 
    					\inferrule*[right= T-Var \ ]
    					{ }
  					{\Gamma^{1} \ \Delta \ \vdash x : \mkterm{Nat} \ \dashv \ \Gamma^{1} \ \Delta} 
  					\ 
  					\inferrule*[right= T-Nat \ ]
    					{ }
  					{\Gamma^{1} \ \Delta \ \vdash 4 : \mkterm{Nat} \ \dashv \ \Gamma^{1} \ \Delta}  
  				}
  				{\Gamma^{1} \ \Delta \ \vdash \ x \le 4 : \mkterm{Bool} \ \dashv \ \Gamma^{1} \ \Delta}  
  				\ \text{T2}
  		}
  		{\Gamma \ \Delta \ \vdash \ \mkterm{while} \ x \le 4 \ \mkterm{do} \ x := x + 1 : \mkterm{Cmd} \ \dashv \ \Gamma\{z \mapsto \mkterm{Nat}\} \ \Delta}  

		\\
		
  		\inferrule*[left=T2 \ ,right=T-Update \ ]
    					{ 
    						\inferrule*[right= T-Var \ ]
    						{ }
  						{\Gamma^{1} \ \Delta \ \vdash x : \mkterm{Nat} \ \dashv \ \Gamma^{1} \ \Delta}
  						\
  						\text{T3}
  					}
  					{\Gamma^{1} \ \Delta \ \vdash x := x + 1 : \mkterm{Nat} \ \dashv \ \Gamma^{1} \ \Delta}

		\\
		\inferrule*[left=T3 \ ,right= T-Var]
	    					{ 
							\inferrule*[right= T-Var \ ]
    							{ }
  							{\Gamma^{1} \ \Delta \ \vdash x : \mkterm{Nat} \ \dashv \ \Gamma^{1} \ \Delta}
							\
  							\inferrule*[right= T-Nat \ ]
    							{ }
  							{\Gamma^{1} \ \Delta \ \vdash 1 : \mkterm{Nat} \ \dashv \ \Gamma^{1} \ \Delta}	    					
	    					}
  						{\Gamma^{1} \ \Delta \ \vdash x + 1 : \mkterm{Nat} \ \dashv \ \Gamma^{1} \ \Delta}
  						
		\\

  \end{mathpar}
  \begin{flushleft}
  $\Gamma = \varnothing$ \\
  $\Gamma^{1} = \{x \mapsto \mkterm{Nat}\}$\\
  $\Delta = \varnothing$
  \end{flushleft}

  \caption{Typing example 2}
  \label{fig:typingexample2}
\end{figure}

\begin{figure}
  \centering
  \begin{mathpar}  
  \inferrule*[right=T-Begin \ ]{
	\text{T1}
	\\ 
	\text{T2}
	\\
	\text{T3}
}{
	\Gamma \ \Delta \ \vdash \ \mkterm{begin} \ Dv \ Dp \ S \ \mkterm{end} : \mkterm{Cmd} \ \dashv \ \Gamma \ \Delta 
}

\\ \\
\inferrule*[left=T1 \ ,right=T-Seq \ ]{
	\inferrule*[right=T-Assign \ ]{
		\inferrule*[right=T-Nat]{ }{
		\Gamma \ \Delta \ \vdash \ 2 : \mkterm{Nat}
	}
	}{
		\Gamma \ \Delta \ \vdash \ \mkterm{var} \ \mkterm{Nat} \ x \ := \ 2 : \mkterm{Cmd} \ \dashv \ \Gamma^{1} \ \Delta
	}
	\\
	\text{T4}
}{
	\Gamma \ \Delta \ \vdash \ \mkterm{var} \ \mkterm{Nat} \ x \ := \ 2 ; \mkterm{var} \ \mkterm{Bool} \ y \ := \ \mkterm{true} : \mkterm{Cmd} \ \dashv \ \Gamma^{2} \ \Delta 
}

\\ \\

\inferrule*[left=T4 \ ,right=T-Assign \ ]{
		\inferrule*[right=T-True]{ }{
		\Gamma^{1} \ \Delta \ \vdash \ \mkterm{true} : \mkterm{Bool}
	}
	}{
		\Gamma^{1} \ \Delta \ \vdash \ \mkterm{var} \ \mkterm{Bool} \ y \ := \ \mkterm{true} : \mkterm{Cmd} \ \dashv \ \Gamma^{2} \ \Delta
	}
	
\\
\inferrule*[left=T2 \ ,right=T-Proc \ ]{
	\inferrule*[right=T-Assign \ ]{
		\inferrule*[right=T-Nat]{ }{
		\Gamma^{2} \ \Delta \ \vdash \ 1 : \mkterm{Nat}
	  }
	}{
		\Gamma^{2} \ \Delta \ \vdash \ \mkterm{var} \ \mkterm{Nat} \ y \ := \ 1 : \mkterm{Cmd} \ \dashv \ \Gamma^{3} \ \Delta 
	}
}{
	\Gamma^{2} \ \Delta \ \vdash \ \mkterm{proc} \ q \ \mkterm{is} \ \mkterm{var} \ \mkterm{Nat} \ y \ := \ 1 : \mkterm{Cmd} \ \dashv \ \Gamma^{2} \ \Delta^{1} 
}
\\ \\
\inferrule*[left=T3 \ ,right=T-Seq \ ]{
	\inferrule*[right=T-Call \ ]
	{ }
	{
		\Gamma^{2} \ \Delta^{1} \ \vdash \ \mkterm{call} \ q : \mkterm{Cmd} \ \dashv \ \Gamma^{4} \ \Delta^{1}}
	\\
	\text{T5}
}
{
	\Gamma^{2} \ \Delta^{1} \ \vdash \ \mkterm{call} \ q ; x \ := \ y : \mkterm{Cmd} \ \dashv \ \Gamma^{4} \ \Delta^{1}
}
\\ \\
\inferrule*[left=T5 \ ,right=T-Update \ ]
	{
		\inferrule*[right=T-Var]{ }{
		\Gamma^{4} \ \Delta \ \vdash \ x : \mkterm{Nat}
	}
		\\ 
		\inferrule*[right=T-Var]{ }{
		\Gamma^{4} \ \Delta \ \vdash \ y : \mkterm{Nat}
	}
	}{
		\Gamma^{4} \ \Delta \ \vdash \ x \ := \ y : \mkterm{Cmd} \ \dashv \ \Gamma^{4} \ \Delta 
	}
  \end{mathpar}
  
  \begin{flushleft}
  $D_v = \mkterm{var} \ \mkterm{Nat} \ x := 2; \mkterm{var} \ \mkterm{Bool} \ y := \mkterm{true}$ \\
  $D_v = \mkterm{proc} \ p \ \mkterm{is} \ \mkterm{var} \ \mkterm{Nat} \ y := 1$ \\
  $S = \mkterm{call} \ p ; x + y$
  \end{flushleft}
  
  \begin{flushleft}
  $\Gamma = \varnothing$ \\
  
  $\Gamma^{1} = \{x \mapsto \mkterm{Nat}\}$ \\
  $\Gamma^{2} = \{x \mapsto \mkterm{Nat},y \mapsto \mkterm{Bool}\}$ \\
  $\Gamma^{3} = \{y \mapsto \mkterm{Nat}\}$ \\
  $\Gamma^{4} = \{x \mapsto \mkterm{Nat},y \mapsto \mkterm{Bool}, y \mapsto \mkterm{Nat}\}$ \\
  $\Delta = \varnothing$
  $\Delta^{1} = \{p \mapsto \Gamma^{3}\}$
  \end{flushleft}
  
  \caption{Typing example 3}
  \label{fig:typingexample3}
\end{figure}

\begin{figure}
  \centering
  \begin{mathpar}  
\inferrule*[right=T-Seq \ ]{
\inferrule*[right=T-Assign \ ]{
		\inferrule*[right=T-Nat]{ }{
		\Gamma \ \Delta \ \vdash \ 1 : \mkterm{Nat}
	}
	}{
		\Gamma \ \Delta \ \vdash \ \mkterm{var} \ \mkterm{Nat} \ y \ := \ 1 : \mkterm{Cmd} \ \dashv \ \Gamma^{1} \ \Delta
	}
	\\
	\text{T1}
}{
	\Gamma \ \Delta \ \vdash \ \mkterm{var} \ \mkterm{Nat} \ y \ := \ 1; \mkterm{begin} \ D_v \ D_p \ S : \mkterm{Cmd} \ \dashv \ \Gamma^{1} \ \Delta 
}

\inferrule*[left=T1 \ ,right=T-Begin]{
		\text{T2}
		\\
		\inferrule*[right=T-Empty]{ }{
		\Gamma \ \Delta \ \vdash \ \epsilon : \mkterm{Cmd}
	}
		\\
		\text{T3}
	}{
		\Gamma^{1} \ \Delta \ \vdash \ \mkterm{begin} \ Dv \ D_p \ S \ \mkterm{end} : \mkterm{Cmd} \ \dashv \ \Gamma^{3} \ \Delta 
	}

\inferrule*[left=T2 \ ,right=T-Seq]{
	\inferrule*[right=T-Assign]{
		\inferrule*[right=T-Nat]{ }{
		\Gamma^{1} \ \Delta \ \vdash \ 2 : \mkterm{Nat}
	}
	}{
		\Gamma^{1} \ \Delta \ \vdash \ \mkterm{var} \ \mkterm{Nat} \ x \ := \ 2 : \mkterm{Cmd} \ \dashv \ \Gamma^{2} \ \Delta
	}
	\\
	\text{T4}
}{
	\Gamma^{1} \ \Delta \ \vdash \ \mkterm{var} \ \mkterm{Nat} \ x \ := \ 2 ; \mkterm{var} \ \mkterm{Bool} \ y \ := \ \mkterm{true} : \mkterm{Cmd} \ \dashv \ \Gamma^{3} \ \Delta 
}
\\ \\
\inferrule*[left=T4 \ ,right=T-Assign]{
	\inferrule*[right=T-True]{ }{
		\Gamma^{2} \ \Delta \ \vdash \ \mkterm{true} : \mkterm{Bool}
	}
	}{
		\Gamma^{2} \ \Delta \ \vdash \ \mkterm{var} \ \mkterm{Bool} \ y \ := \ \mkterm{true} : \mkterm{Cmd} \ \dashv \ \Gamma^{3} \ \Delta
	}
\\ \\
\inferrule*[left=T3 \ ,right=T-Update]{
	\inferrule*[right=T-Var]{ }{
			\Gamma^{3} \ \Delta \ \vdash \ x : \mkterm{Nat}
		}
	\\
	\text{T5}
}{
	\Gamma^{3} \ \Delta \ \vdash \ x \ := \ x \ + \ y : \mkterm{Cmd} \ \dashv \ \Gamma^{3} \ \Delta 
}
\\ \\
\inferrule*[left=T5,right=T-Add]{
		\inferrule*[right=T-Var]{ }{
			\Gamma^{3} \ \Delta \ \vdash \ x : \mkterm{Nat}
		}
		\\
		\Gamma^{3} \ \Delta \ \vdash \ y : \mkterm{Bool}
	}{
		\Gamma^{3} \ \Delta \ \vdash \ x \ + \ y : \mkterm{Nat} \ \dashv \ \Gamma^{3} \ \Delta 
	}
  \end{mathpar}
  
  \begin{flushleft}
  $\Gamma = \varnothing$ \\
  $\Gamma^{1} = \{y \mapsto \mkterm{Nat}\}$ \\
  $\Gamma^{2} = \{y \mapsto \mkterm{Nat},x \mapsto \mkterm{Nat}\}$ \\
  $\Gamma^{3} = \{y \mapsto \mkterm{Nat},x \mapsto \mkterm{Nat},y \mapsto \mkterm{Bool}\}$\\
  $\Delta = \varnothing$
  \end{flushleft}
  \caption{Typing example 4}
  \label{fig:typingexample4}
\end{figure}

\clearpage

\section{Testing the While formalization}
\label{sec:racket}
Formally defining a programming language is important since such definition can help detect design errors in the language and interpret and evaluate programs. Since producing derivations of executions or of typing is tedious and error-prone, implementing the reduction rules and the type system is crucial to avoid the above mentioned difficulties but may be very time consuming. 

In this section we introduce the Racket language, a programming language that supports other programming languages.

\subsection{Racket}
Racket is a programming language in the Lisp family, meaning that while it can be used to create solutions like any conventional programming language, it also allows a language-oriented programming, i.e., allows creating new programming languages. To support this feature, Racket provides building blocks for protection mechanisms, which allows the programmers to protect individual components of the language from their clients, and the internalization of extra-linguistic mechanisms, such as project contexts and the delegation of program execution and inspection to external agents, by converting them into linguistic constructs, preventing programmers to resort to mechanism outside Racket \cite{manifesto}.

\subsection{PLT Redex}
PLT Redex is a domain-specific language embedded in Racket that allows programmers to formalize and debug programming languages. The modeling of a programming language in Redex is done by writing down the grammar, reductions of the language along with necessary metafunctions. Since Redex is embedded in Racket, programming in Redex is just like programming in Racket, with all of the features and tools available for Racket being also available for Redex, including DrRacket, a integrated development environment for Racket. One of the most interesting advantages of using DrRacket is the automatically generated reduction graphs that allows programmers to visualize reductions step by step. Redex also has other methods of testing, such as pattern matcher (for grammar testing) and judgment-form evaluation (which we use to test the type system) \cite{redex, runresearch}.

PLT Redex is the tool we choose to help us certify our work. To understand it better, we implemented the While language as formalized in this article using PLT Redex \footnote{Available at \url{https://bitbucket.org/cvasconcelos/thesis/src/876fc254db76aca1bb058b7e6ef069ee23c6c237/While/while.rkt}}. We recommend using DrRacket while trying this and other implementations we provide since it is necessary to visualize the generated reduction graphs. 

\begin{figure}
\centering
\includegraphics[scale = 0.5]{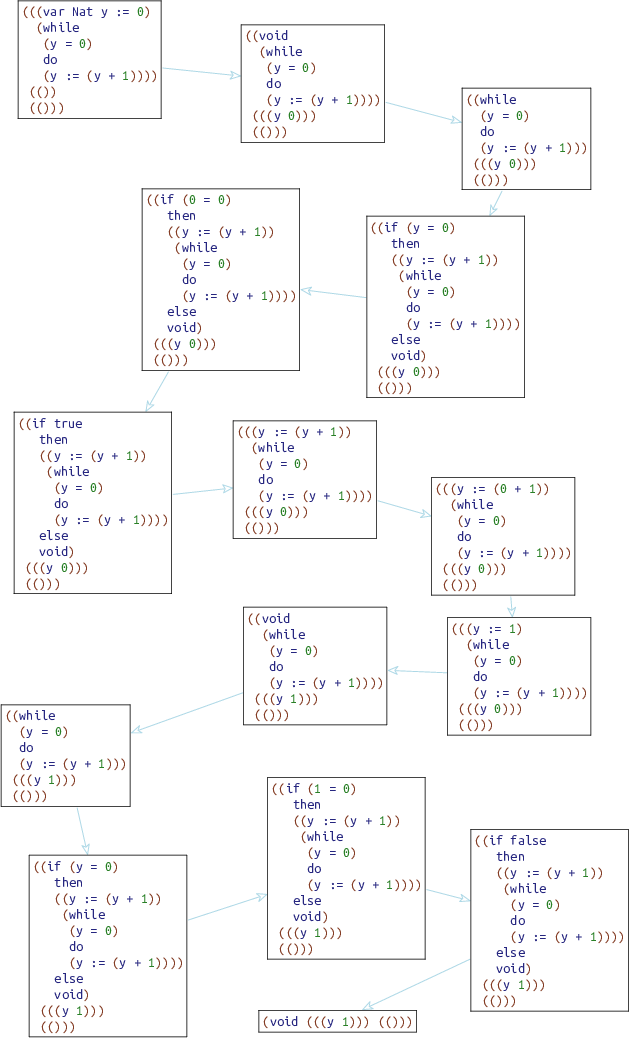} 
\caption{PLT-Redex reduction graph example 1}
\label{fig:redex3}
\end{figure}

\begin{figure}
\centering
\includegraphics[scale = 0.5]{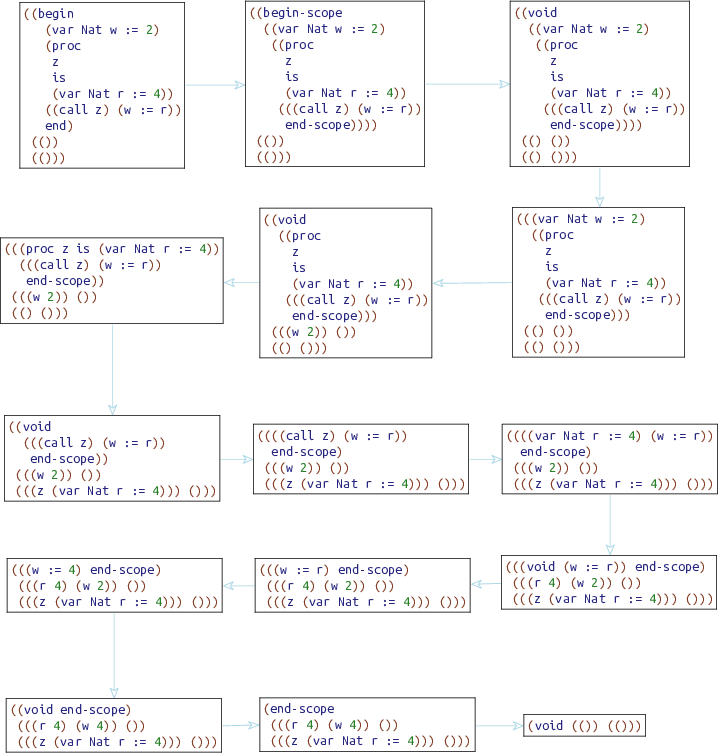} 
\caption{PLT-Redex reduction graph example 2}
\label{fig:redex4}
\end{figure}

\begin{figure}
\centering
\includegraphics[scale = 0.5]{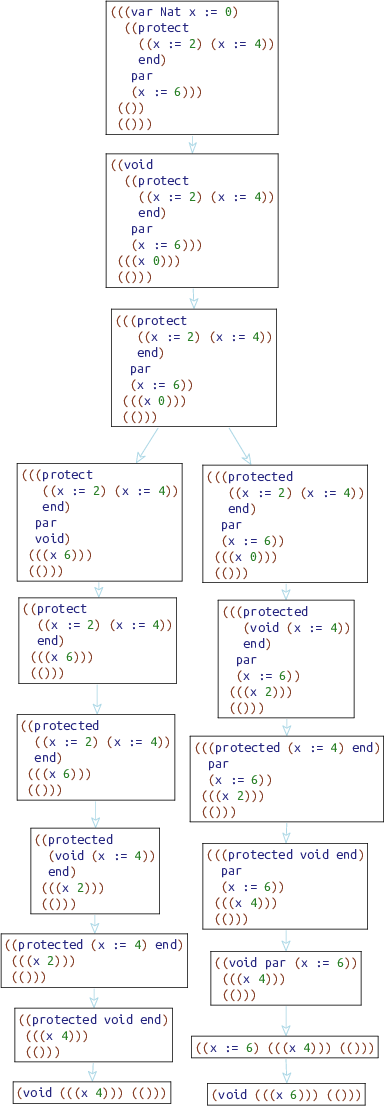} 
\caption{PLT-Redex reduction graph example 3}
\label{fig:redex5}
\end{figure}

Figures \ref{fig:redex3}, \ref{fig:redex4} and \ref{fig:redex5} show the reduction graphs for simple program examples for the While language.

\section{Conclusions and further work}
We present and offer an implementation in Racket \cite{manifesto}, a
programming language that supports other programming languages, of the
language While described in a book by Hanne Riis and Flemming
Nielson~\cite{riisnielsonh.nielsonf.2007}. This implementation
directly represents the original syntax and operational semantics of
While, faithfully following the definitions presented in the book. One
can now automatically build derivations of the possible reductions of
any program, observing its step-by-step execution.

Moreover, we define an original type system for the language While,
which we also implemented in Racket to provide an automatic
type-checking engine.

Future work include stating and proving properties like subject
reduction and type safety, in a system like Why3~\cite{FilliatrePaskevich:why3}.

\printbibliography

\end{document}